\documentclass[aps,prl,preprint,superscriptaddress,showpacs,byrevtex,nofootinbib]{revtex4}
\usepackage{graphicx} 
\usepackage{dcolumn}  
\usepackage{color}
\usepackage{amsmath}

\graphicspath{{ps}}

\newcommand{\hhh}{$\tau^{-}\rightarrow h^{-} h^{+} h^{-}\nu_{\tau}$}
\newcommand{\Kpipi}{$\tau^{-}\rightarrow K^{-} \pi^{+}\pi^{-}\nu_{\tau}$}
\newcommand{\pipipi}{$\tau^-\rightarrow \pi^-\pi^+\pi^-\nu_{\tau}$}
\newcommand{\KKpi}{$\tau^-\rightarrow K^-K^+\pi^-\nu_{\tau}$}
\newcommand{\KKK}{$\tau^-\rightarrow K^-K^+K^-\nu_{\tau}$}
\newcommand{\Kspi}{$\tau^-\rightarrow K_S^0\pi^-\nu_{\tau}$}
\newcommand{\enunu}{$\tau\rightarrow e\overline{\nu}\nu$}
\newcommand{\mununu}{$\tau\rightarrow \mu\overline{\nu}\nu$}
\newcommand{\pipipipizero}{$\tau^-\rightarrow \pi^-\pi^+\pi^-\pi^0\nu_{\tau}$}
\newcommand{\eeqq}{$e^+e^-\rightarrow q\bar{q}$}

\newcommand{\MKpipi}{$M(K\pi\pi)$}
\newcommand{\Mpipipi}{$M(\pi\pi\pi)$}
\newcommand{\MKKpi}{$M(KK\pi)$}
\newcommand{\MKKK}{$M(KKK)$}

\newcommand{\Ks}{\textcolor{black}{$K_S^0$}}
\newcommand{\pizero}{$\pi^0$}
\newcommand{\fbi}{$\rm {fb^{-1}}$}
\newcommand{\plusminus}{$\pm$}

\newcommand{\GeVc}{GeV/$c$}

\newcommand{\GeVcc}{GeV/$c^2$}
\newcommand{\MeVcc}{MeV/$c^2$}
\newcommand{\emuevent}{\{$\,e,\mu\,$\}}
\newcommand{\eetautau}{$e^+e^-\rightarrow \tau^+\tau^-$}
\newcommand{\degree}{$^{\circ}$}
\newcommand{\Br}{$\mathcal{B}$}

\begin{document}

\vspace*{-3\baselineskip}
\resizebox{!}{3cm}{\includegraphics{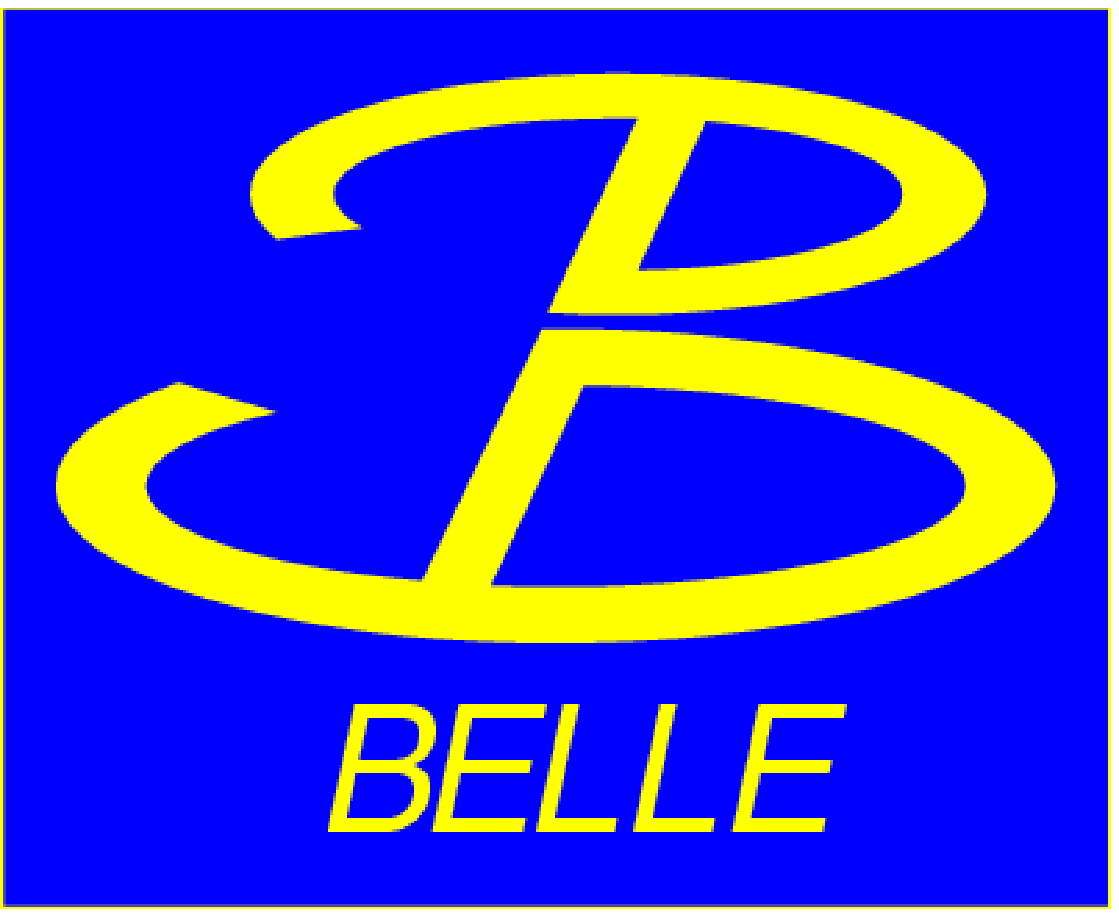}}

\preprint{\vbox{ \hbox{   }
		\hbox{Belle Preprint {\it 2009-23}}
		\hbox{KEK Preprint {\it 2009-26}}
}}

\title{ \quad\\[1.0cm] Measurement of the branching fractions and the invariant mass distributions for {\hhh} decays}


\affiliation{Budker Institute of Nuclear Physics, Novosibirsk}
\affiliation{University of Cincinnati, Cincinnati, Ohio 45221}
\affiliation{Justus-Liebig-Universit\"at Gie\ss{}en, Gie\ss{}en}
\affiliation{The Graduate University for Advanced Studies, Hayama}
\affiliation{Hanyang University, Seoul}
\affiliation{University of Hawaii, Honolulu, Hawaii 96822}
\affiliation{High Energy Accelerator Research Organization (KEK), Tsukuba}
\affiliation{Institute of High Energy Physics, Chinese Academy of Sciences, Beijing}
\affiliation{Institute of High Energy Physics, Vienna}
\affiliation{Institute of High Energy Physics, Protvino}
\affiliation{Institute for Theoretical and Experimental Physics, Moscow}
\affiliation{J. Stefan Institute, Ljubljana}
\affiliation{Kanagawa University, Yokohama}
\affiliation{Korea University, Seoul}
\affiliation{Kyungpook National University, Taegu}
\affiliation{\'Ecole Polytechnique F\'ed\'erale de Lausanne (EPFL), Lausanne}
\affiliation{Faculty of Mathematics and Physics, University of Ljubljana, Ljubljana}
\affiliation{University of Maribor, Maribor}
\affiliation{Max-Planck-Institut f\"ur Physik, M\"unchen}
\affiliation{University of Melbourne, School of Physics, Victoria 3010}
\affiliation{Nagoya University, Nagoya}
\affiliation{Nara Women's University, Nara}
\affiliation{National Central University, Chung-li}
\affiliation{National United University, Miao Li}
\affiliation{Department of Physics, National Taiwan University, Taipei}
\affiliation{H. Niewodniczanski Institute of Nuclear Physics, Krakow}
\affiliation{Nippon Dental University, Niigata}
\affiliation{Niigata University, Niigata}
\affiliation{University of Nova Gorica, Nova Gorica}
\affiliation{Novosibirsk State University, Novosibirsk}
\affiliation{Osaka City University, Osaka}
\affiliation{Panjab University, Chandigarh}
\affiliation{University of Science and Technology of China, Hefei}
\affiliation{Seoul National University, Seoul}
\affiliation{Shinshu University, Nagano}
\affiliation{Sungkyunkwan University, Suwon}
\affiliation{School of Physics, University of Sydney, NSW 2006}
\affiliation{Tata Institute of Fundamental Research, Mumbai}
\affiliation{Excellence Cluster Universe, Technische Universit\"at M\"unchen, Garching}
\affiliation{Toho University, Funabashi}
\affiliation{Tohoku Gakuin University, Tagajo}
\affiliation{Department of Physics, University of Tokyo, Tokyo}
\affiliation{Tokyo Metropolitan University, Tokyo}
\affiliation{Tokyo University of Agriculture and Technology, Tokyo}
\affiliation{IPNAS, Virginia Polytechnic Institute and State University, Blacksburg, Virginia 24061}
\affiliation{Yonsei University, Seoul}
  \author{M.~J.~Lee}\affiliation{Seoul National University, Seoul} 
  \author{H.~Aihara}\affiliation{Department of Physics, University of Tokyo, Tokyo} 
  \author{K.~Arinstein}\affiliation{Budker Institute of Nuclear Physics, Novosibirsk}\affiliation{Novosibirsk State University, Novosibirsk} 
  \author{V.~Aulchenko}\affiliation{Budker Institute of Nuclear Physics, Novosibirsk}\affiliation{Novosibirsk State University, Novosibirsk} 
  \author{T.~Aushev}\affiliation{\'Ecole Polytechnique F\'ed\'erale de Lausanne (EPFL), Lausanne}\affiliation{Institute for Theoretical and Experimental Physics, Moscow} 
  \author{A.~M.~Bakich}\affiliation{School of Physics, University of Sydney, NSW 2006} 
  \author{E.~Barberio}\affiliation{University of Melbourne, School of Physics, Victoria 3010} 
  \author{A.~Bay}\affiliation{\'Ecole Polytechnique F\'ed\'erale de Lausanne (EPFL), Lausanne} 
  \author{K.~Belous}\affiliation{Institute of High Energy Physics, Protvino} 
  \author{V.~Bhardwaj}\affiliation{Panjab University, Chandigarh} 
% ...
\collaboration{The Belle Collaboration}


\begin{abstract}

We present a study of 
{\pipipi}, {\Kpipi}, {\KKpi}, and {\KKK} decays using 
a 666 {\fbi} data sample 
collected with the Belle detector at the KEKB asymmetric-energy 
$e^+ e^-$ collider 
at and near
a center-of-mass energy of 10.58 GeV.
The branching fractions are measured to be: 
{\Br}({\pipipi})$ = (8.42 \pm 0.00 ^{+0.26} _{-0.25}) \times 10^{-2}$, 
{\Br}({\Kpipi})$ = (3.30 \pm 0.01 ^{+0.16} _{-0.17}) \times 10^{-3}$, 
{\Br}({\KKpi})$ = (1.55 \pm 0.01 ^{+0.06} _{-0.05}) \times 10^{-3}$, and
{\Br}({\KKK})$ = (3.29 \pm 0.17 ^{+0.19} _{-0.20}) \times 10^{-5}$, 
where the first uncertainty is statistical and the second is
systematic. 
These branching fractions do not include contributions 
from modes in which a $\pi^+ \pi^-$ pair originates from a {\Ks} decay. 
We also present the unfolded invariant mass distributions for these decays. 

\end{abstract}

\pacs{13.35.Dx, 12.15.Hh, 14.60.Fg}

\maketitle


{\renewcommand{\thefootnote}{\fnsymbol{footnote}}}
\setcounter{footnote}{0}

\section{Introduction}
\label{Section:Introduction}
\begin{figure}[htb]
\begin{center}
\includegraphics[width=0.6\textwidth]
{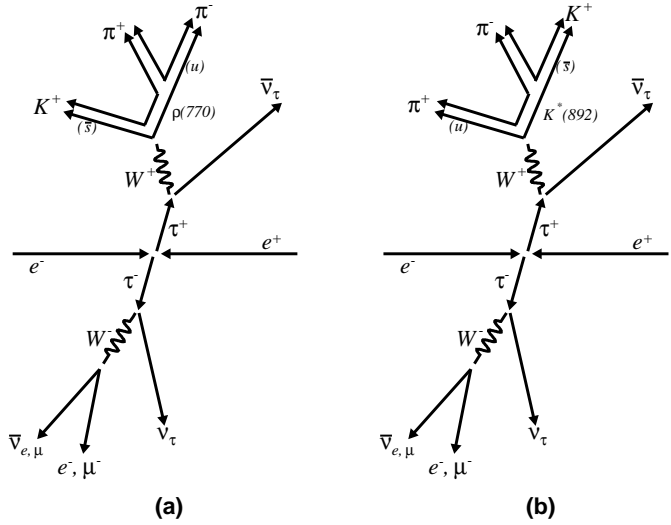}
\caption[Event shape]{A schematic view of {\eetautau} in 
the center-of-mass system, where one $\tau$ ($\tau^+$) decays 
to the signal mode ($\tau^+\rightarrow K^+\pi^-\pi^+\nu_{\tau}$) 
and the other $\tau$ ($\tau^-$) decays to a leptonic  
mode ($\tau^-\rightarrow e^-\nu_{\tau}\overline{\nu}_{e}$ or 
$\tau^-\rightarrow \mu^-\nu_{\tau}\overline{\nu}_{\mu}$). 
Each figure shows one of two possible intermediate resonant 
states in the signal mode. }
\label{EventShape}
\end{center}
\end{figure}

The decays of the $\tau$ lepton into three pseudoscalar particles can 
provide information on hadronic form factors,
$K^*$ resonance spectroscopy, and the Wess-Zumino anomaly
\cite{Citation:WessZumino},
and also can be used for studies of CP violation in the leptonic sector \cite{Citation:CPV}.
By studying decays into final states that contain one or three kaons,
it is possible to extract strange spectral functions, which can be used for a direct determination of 
the strange quark mass and the Cabibbo-Kobayashi-Maskawa (CKM) 
matrix element $|V_{us}|$~\cite{Citation:VusMs,Citation:ALEPH,Citation:OPAL}.
These three hadron decays have been 
studied since the initial discovery of the $\tau$ lepton, 
but only the decay to three charged pions has been closely investigated.
Even for this mode, measurements of its branching fraction have only been provided by the 
CLEO~\cite{Citation:CLEO} and BaBar~\cite{Citation:BABAR} groups.
For all other modes, there are still many unresolved problems.     
For example, the branching fraction for {\Kpipi} decay recently
measured by the BaBar collaboration~\cite{Citation:BABAR} 
is significantly lower
than the values from most 
previous measurements~\cite{Citation:PDG};
there is large scatter in the measured
central values, 
which is reflected in the large scale factor (2.1) 
applied by the Particle Data Group when forming their average~\cite{Citation:PDG}.
Intermediate two- and three-body resonance states in  this decay 
have been previously studied but only with limited statistical precision, see, e.g.,
Refs~\cite{Citation:ALEPHResonance,Citation:OPALResonance, 
Citation:CLEOResonance}. 
There is evidence for quite rich dynamics, with strong signals for the $K_1(1270)$,
$K_1(1400)$ and $K^*(1410)$ resonances in the three-body final state;
other resonances can be seen in various two-body sub-systems.   
Theoretically, two intermediate resonances, the $\rho(770)^0$ and 
$K^{*}(892)^0$ (or its excitations), are expected to contribute to 
$\tau$ decay into $K^-\pi^+\pi^-\nu_{\tau}$:
$\tau^{-}\rightarrow K^{-} \rho(770)^0 (\rightarrow \pi^{+} \pi^{-}) \nu_{\tau}$ and 
$\tau^{-}\rightarrow K^*(892)^0 (\rightarrow K^{-} \pi^{+}) \pi^{-} \nu_{\tau}$ \cite{Citation:Resonance}.
These are shown in Figs~\ref{EventShape}(a) and (b), respectively. 
 
Here we present new measurements of the branching fractions
for {\pipipi}, {\Kpipi}, {\KKpi}, and {\KKK} decays. (Unless otherwise 
specified, the charge-conjugate decay is also implied throughout this
paper.) 
Events in which a $\pi^+ \pi^-$ pair is consistent with a {\Ks} decay are excluded. 
Because of  particle misidentification, measurements of all
four of the above modes are correlated and, thus, are considered simultaneously. 

We use a data sample of  
666 {\fbi} collected with the Belle detector at the KEKB 
asymmetric-energy $e^+e^-$ collider~\cite{Citation:KEKB}
on the $\Upsilon$(4S) resonance, 10.58 GeV, 
and 60 MeV below it (off-resonance). 
This sample contains 
$6.12\times10^8$ produced $\tau$-pairs.
The Belle detector is a large-solid-angle magnetic
spectrometer that consists of a silicon vertex detector (SVD),
a 50-layer central drift chamber (CDC), an array of
aerogel threshold Cherenkov counters (ACC), 
a barrel-like arrangement of time-of-flight
scintillation counters (TOF), and an electromagnetic calorimeter (ECL)
comprised of CsI(Tl) crystals located inside
a superconducting solenoid coil that provides a 1.5~T
magnetic field.  An iron flux return located outside 
the coil (KLM) is instrumented to detect $K_L^0$ mesons and to identify
muons.  The detector
is described in detail elsewhere~\cite{Citation:Belle}.

\section{Selection of events}
\label{Section:SelectionOfEvent}

Three-prong decays of the $\tau$ are selected from 
candidate $\tau^+\tau^-$ pair events as follows. 
We use events where the number of charged tracks is four 
and the sum of their charges is zero; 
the transverse momentum of each track in the laboratory frame is required to 
be larger than 0.1 {\GeVc},  
and the tracks should extrapolate back to the interaction point 
within {\plusminus}5 cm along the beam direction
and {\plusminus}1 cm in the transverse plane.
The sum of the reconstructed momenta in the center-of-mass (CM) 
frame is required to be less than 10 {\GeVc}, 
and the sum of energies deposited in the calorimeter is 
required to be less than 10 GeV.
The largest of the four tracks' transverse momenta is required to be greater than 0.5 {\GeVc}
to reject two-photon events, which typically have low transverse momentum tracks.
To reject beam-related background, 
we require the position of the reconstructed event vertex
to be less than 3 cm from the interaction point along the
beam direction, and separated by less than 0.5 cm in the
transverse plane.
The missing mass, 
$M_{\rm {miss}}^2 = 
({\it p}_{\rm {init}} - \sum_{\rm {tr}}{\it p_{\rm {tr}}} - 
\sum_{\gamma}{\it p_{\gamma}})^2$,
and the polar angle of the missing momentum in the CM frame 
are efficient variables for rejecting two-photon and Bhabha backgrounds.
In the definition of the missing mass, 
${\it p_{\rm {tr}}}$ and ${\it p_{\gamma}}$ are the four-momenta of 
measured tracks and photons, respectively,
and ${\it p}_{\rm {init}}$ is the initial CM frame four-momentum 
of the $e^+e^-$ beams. 
We require that the missing mass be larger than 1 {\GeVcc} and 
less than 7 {\GeVcc}, 
and that the polar angle with respect to the beam direction 
in the CM frame be larger than 
30{\degree} and less than 150{\degree}. 
Detailed information on the $\tau^+\tau^-$ pair selection and 
background suppression using missing mass 
can be found in~\cite{Citation:pipizero}.

\begin{figure}[htb]
\begin{center}
\includegraphics[width=0.9\textwidth]
{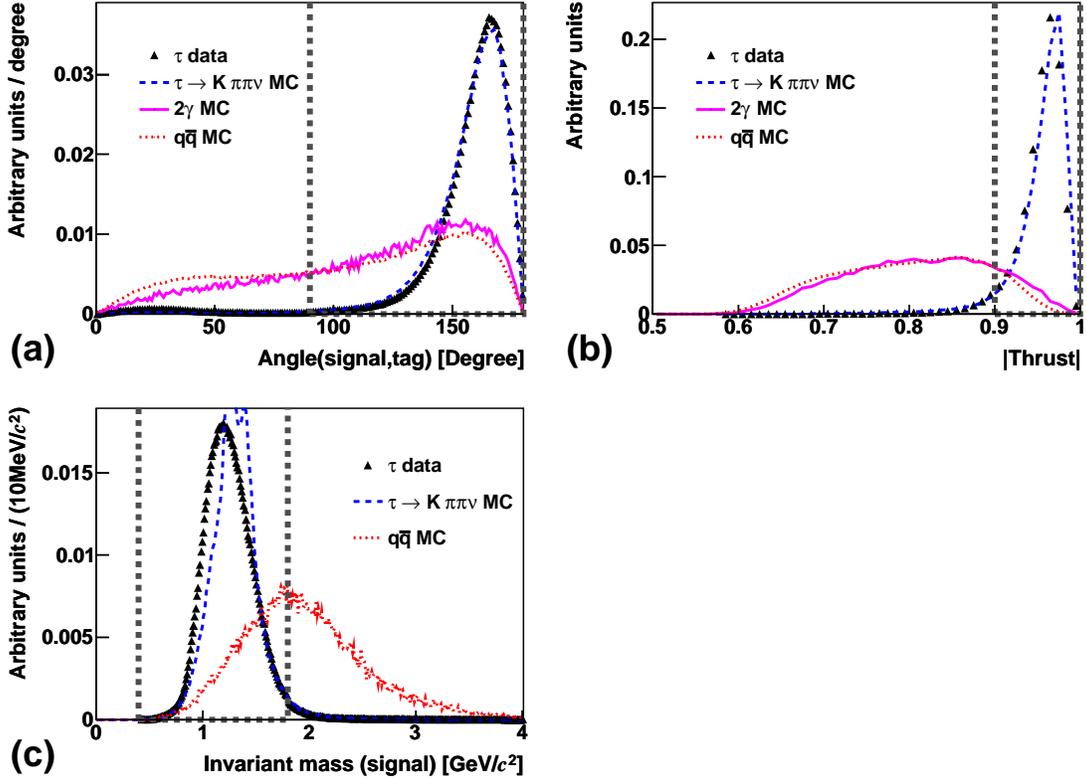}
\caption{Distributions of variables used in the event selection.  
(a) The angle between the total momentum of the hadronic system on 
the signal-side and the lepton momentum on the tag-side in the CM system.
(b) The magnitude of the thrust.
(c) The invariant mass of the hadronic system.
The triangular points show data, the dashed histograms show the 
{\Kpipi} signal MC, the solid histograms show the two-photon background, 
and the dotted histograms show the {\eeqq} background. 
The dotted vertical lines show the boundaries used to select the signal candidates.
All samples are normalized to the same number of events.}
\label{CutStatus}
\end{center}
\end{figure}

Particle identification 
is used to select $\tau$ events that contain one lepton 
(electron or muon) and three hadrons (pions or kaons). 
The magnitude of thrust is evaluated and is required to be larger 
than 0.9 to suppress two-photon and {\eeqq} backgrounds, 
where the thrust is defined by the maximum of 
$(\sum_i | \hat{n} \cdot \vec{p_i}|) / (\sum_i |\vec{p_i}|)$.
Here $\vec{p_i}$ is the momentum of the $i$-th track and 
$\hat{n}$ is the unit vector in the direction of the thrust 
axis -- the direction maximizing the thrust. 
We require that the angle between the total momentum of the hadrons 
and the lepton momentum in the CM system be larger than 90{\degree},  
since the tag-side lepton and signal-side hadrons 
usually
lie in opposite hemispheres: 
the so-called 1--3 prong configuration, 
due to the large transverse momenta of $\tau$ leptons. 
The invariant mass of charged tracks and gamma clusters on each side 
is required to be less than the $\tau$ mass. 
Finally, we require that there be no {\Ks}, {\pizero}, 
or energetic $\gamma$ on the signal side,
where the selection criteria for these particles are described below.
Figure~\ref{CutStatus} shows 
distributions of the variables
used to select the 1--3 prong sample, 
where in each case, the selection value is indicated by a vertical line in the figure.
The surviving events are candidates 
for $\tau^-\rightarrow h_1^-h_2^+h_3^-\nu_{\tau}$,
where $h_{1,2,3}$ is either a pion or a kaon candidate. 
At this stage, the efficiencies for reconstructing 
{\pipipi}, {\Kpipi}, {\KKpi}, and {\KKK} decays 
as {\hhh}  
are $26.7\%$, $27.5\%$, $27.2\%$, and $24.8\%$, respectively, 
while the reconstruction efficiencies of the dominant background modes, {\pipipipizero}, {\Kspi}, 
and {\eeqq}, 
are 6.0\%, 1.8\%, and 0.004\%, respectively. 
The fractions of {\pipipipizero}, {\Kspi}, {\eeqq} and two-photon backgrounds that are reconstructed as {\hhh} are
$6.47\%$, $0.34\%$, $0.35\%$ and $0.05\%$, 
respectively;
the two-photon background contribution is negligible.
 
\begin{figure}[htb]
\begin{center}
\includegraphics[width=0.6\textwidth]
{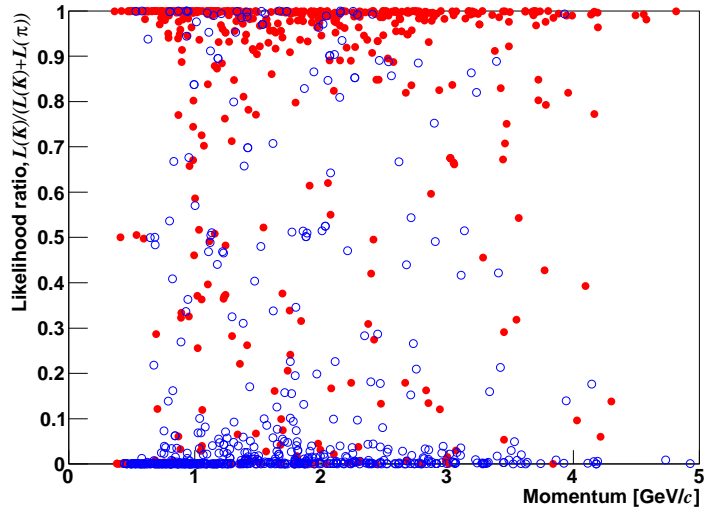}
\caption[Kaon ID likelihood ratio] {
Likelihood ratio for kaon 
identification, $L(K)/(L(K)+L(\pi))$, versus particle momentum.
The filled and open circles correspond to kinematically identified
kaons and pions, respectively.} 
\label{KidLikelihoodRatio}
\end{center}
\end{figure}

An important
issue for this analysis is the separation 
of kaons and pions. 
In Belle, $dE/dx$ information from the CDC, 
hit information from the ACC, and time-of-flight measurements from the TOF system
are used to construct a likelihood for the kaon (pion) hypothesis, 
$L(K)$ ($L(\pi)$). 
Figure \ref{KidLikelihoodRatio} shows the kaon likelihood ratio, 
$L(K)/(L(K)+L(\pi))$, as a function of momenta. 
Clean kaon-pion separation is evident over the 
entire momentum range relevant for this analysis, 
where the track momentum ranges from 0.1 {\GeVc} 
to $\sim$5 {\GeVc}, and the average momentum is $\sim$1.3 {\GeVc}.
As discussed below, we choose a relatively stringent particle identification 
(PID) requirement for kaons ($L(K)/(L(K)+L(\pi)) > 0.9$) and a  
less restrictive one for pions ($L(K)/(L(K)+L(\pi)) < 0.9$).  
The kaon identification selection has an
efficiency of $\sim$73\% for kaons and rejects $\sim$95\% of pions.
The kaon and pion identification efficiencies are calibrated 
using kaon and pion tracks in data from kinematically identified $D^{*+}$ decays,
$D^{*+} \rightarrow D^0 (\rightarrow K^- \pi^+) \pi^+$. 
We evaluate the efficiencies and fake rates for this calibration
sample and compare them to Monte Carlo (MC) expectations.
From this comparison, we obtain a correction table as a 
function of track momenta ($p_{lab}$) and polar angles ($\theta_{lab}$), 
and apply it to the MC.
The accuracy of the  correction factor,
which is the source of the systematic uncertainty of the evaluation of
the branching fraction and the mass spectra, 
is limited by the statistical uncertainties  
of kaon and pion samples from $D^{*+}$ decays 
in the certain $p_{lab}$ and $\theta_{lab}$ bins,
and the uncertainty of the $D^{*+}$ signal extraction. 

\begin{figure}[htb]
\begin{center}
\includegraphics[width=0.6\textwidth]
{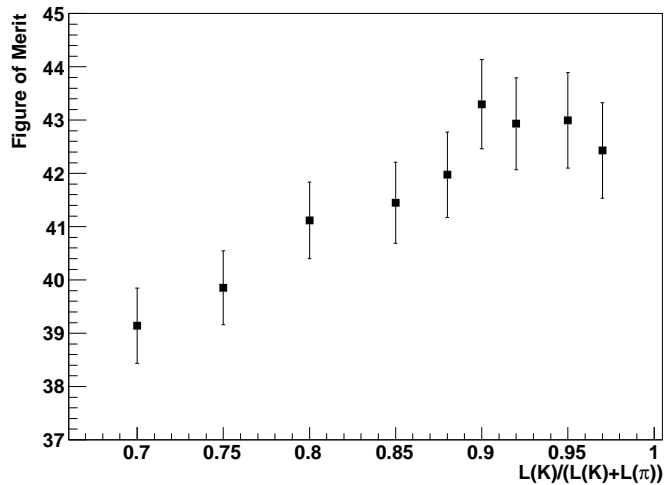}
\caption[PID study result]
{Figure-of-merit 
as a function of the likelihood ratio $L(K)/(L(K)+L(\pi))$ requirement, 
used to separate the kaon and pion hypotheses.
The uncertainties are evaluated 
from the comparison of the efficiencies and fake rates of 
kaon identification with those of the  
$D^{*+} \rightarrow D^0 (\rightarrow K^- \pi^+) \pi^+$~ control sample.}
\label{PIDstudy1}
\end{center}
\end{figure}

The kaon identification criterion is determined by 
maximizing a figure-of-merit (FOM), 
where the requirement on the kaon identification likelihood ratio is varied.
The figure-of-merit we used is defined as:
\begin{equation}
FOM = \frac{S}{\sqrt{S+N}}~,
\end{equation}
where $S$ is the number of signal events 
and $N$ is the number of cross-feed background events. 
For the {\Kpipi} signal ($S$), 
the cross-feed background component ($N$) includes contributions from {\pipipi} and {\KKpi} decays.
The result of FOM optimization for the {\Kpipi} decay is shown 
in Fig.~\ref{PIDstudy1}, where one can see that FOM is indeed
maximal for the particle identification criteria used in this
analysis. 
The same FOM analysis was performed for the {\pipipi} and {\KKpi} 
decay modes and resulted in similar particle 
identification criteria.
For electron identification, the likelihood variable is calculated based on 
the track extrapolation to the ECL, the ratio of the energy deposited 
in the ECL to the momentum measured in the CDC,
the measured $dE/dx$ in the CDC, the shower shape in the ECL, 
and light yield in the ACC \cite{Citation:eid}.
The extrapolation of tracks from the CDC and SVD to the KLM is used 
to construct the likelihood variable for muon 
identification~\cite{Citation:muid}.
The efficiencies and systematic uncertainties for lepton identification 
are evaluated by using a  
$\gamma\gamma\rightarrow e^+e^-/\mu^+\mu^-$ control sample.

To reduce the feed-down background
and to estimate the remaining  
backgrounds that contain {\Ks} and {\pizero} mesons,
we reconstruct {\Ks} and {\pizero} signals explicitly.
{\Ks} candidates are formed from oppositely charged pairs of pion tracks with  
invariant mass $M(\pi\pi)$ within 
{\plusminus}13.5 {\MeVcc} ($\pm5\sigma$) of the {\Ks} mass.
To improve the {\Ks} purity, 
the point of closest approach to the interaction point along 
the extrapolation of each track
is required to be larger than 0.3 cm 
in the plane transverse to the beam direction.
The azimuthal angle between the momentum vector and the decay vertex 
vector of the reconstructed {\Ks} is required to be less than 0.1 rad. 
The distance between the two daughter pion tracks at their 
point of closest approach is required to be less than 1.8 cm, and the 
flight length of the {\Ks} 
in the plane perpendicular to the beam direction
is required
to be larger than 0.08 cm. 
Finally, we require the invariant mass of each {\Ks} 
candidate to be inconsistent with that of a $\Lambda$, $\bar{\Lambda}$, 
and a photon conversion, 
where the daughter tracks are assumed to be 
pions and protons (antiprotons) or electrons and positrons as appropriate.
Candidate {\pizero}'s 
are reconstructed from pairs of photon clusters. 
The energy of each photon is required to exceed 50 MeV for  
candidates in the barrel part of the calorimeter, and 100 MeV for 
candidates in the endcap part of the calorimeter. 
We also require the invariant mass of the two photons,
$M(\gamma\gamma)$, to be within {\plusminus} 16 {\MeVcc} of the {\pizero} mass. 
Events containing one 
or more reconstructed {\Ks} or {\pizero} are rejected.
In addition, in order to reject events containing {\pizero}'s,
we require that there be no photon with energy larger than 0.3 GeV. 

\section{Efficiency estimation}
\begin{figure}[htb]
\begin{center}
\includegraphics [width=0.6\textwidth]
{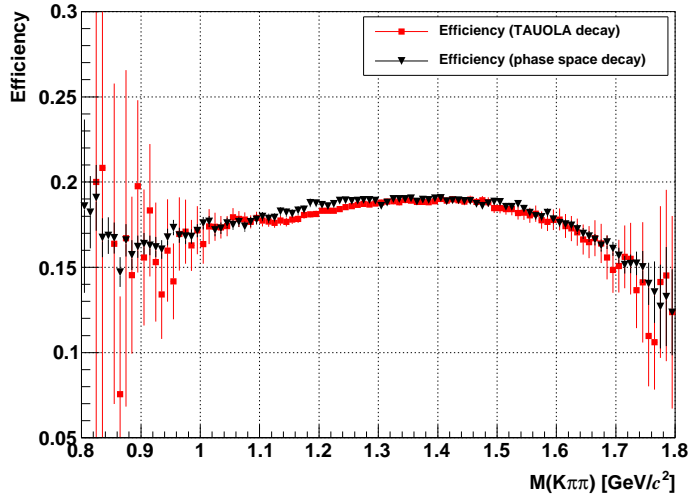}
\caption[Efficiency of different decay model]{The efficiency for 
the {\Kpipi} decay as a function of {\MKpipi}. Squares and triangular 
points represent the 
TAUOLA
decay model and phase-space decay, respectively.}
\label{EfficiencyCurve}
\end{center}
\end{figure}

To estimate the signal efficiency, 
97 million signal MC events for the {\pipipi} 
decay mode and 5 million signal MC events for each of the {\Kpipi}, {\KKpi}, and {\KKK} modes were generated
using the TAUOLA~\cite{Citation:TAUOLA} based KKMC~\cite{Citation:KKMC} generator.
The detector response for all MC data sets was simulated with 
the GEANT3~\cite{Citation:GEANT} package. 
The TAUOLA generator provides models for {\pipipi},
{\Kpipi}, and {\KKpi} decays, where various intermediate resonances 
are taken into account.
To check whether there is  efficiency bias related to
the specific decay model used, 
events 
with phase-space decay distributions were
generated using the KKMC program.
The efficiencies as functions of the invariant mass 
for both {\Kpipi} decay models are compared in 
Fig.~\ref{EfficiencyCurve}.
The relative difference in efficiency between the 
TAUOLA  
and phase-space decay models is around 1\%.           
Similarly, very small discrepancies between the 
TAUOLA
and phase-space decay model are observed in the {\pipipi} 
and {\KKpi} decays. We used the 
TAUOLA
decay model to evaluate 
the efficiencies 
and their dependencies on the invariant mass 
of {\pipipi}, {\Kpipi}, and {\KKpi} decays.  
For the {\KKK} decay mode, the TAUOLA generator does not provide 
any decay model, therefore the efficiency is 
calculated assuming a phase-space decay.
In practice, the efficiencies are evaluated using a two step procedure.
First, we evaluate the response 
matrix for unfolding the mass spectra
using the TAUOLA decay model.
Next, we recalibrate the efficiencies using the unfolded spectra.
Details of this procedure are described in the next section.

The average efficiencies and the fractions of the cross-feed 
backgrounds for all three-prong decays are summarized in 
Table \ref{EfficiencyTable}. 
The probability of reconstructing {\pipipi}  as {\Kpipi} is relatively low, but due to
the large branching fraction of {\pipipi} decay, 
there is a substantial contamination 
of misidentified {\pipipi} events
in the {\Kpipi} sample.
For a similar reason, the {\KKK} decay mode has  a large 
cross-feed background from {\KKpi}.

\begin{center}
\begin{table}[!hbp]
\caption{Summary of the efficiencies and the fractions of cross-feed backgrounds.
The values in parentheses are the 
differences of the efficiencies or fake rates from those
evaluated directly from the MC with the TAUOLA decay model, 
$(\mathcal{E}^{(2)}_{ij} - \mathcal{E}^{(1)}_{ij})/\mathcal{E}^{(1)}_{ij}$  in percent (see text).
}
\begin{tabular}{l|c|c|c|c}
\hline
 & \multicolumn{4}{c}{Generated decay mode} \\
\cline{2-5} 
Reconstructed 	& $\tau^-\rightarrow$ & $\tau^-\rightarrow$  & 
$\tau^-\rightarrow$ 		&  $\tau^-\rightarrow$ \\
decay mode 	& $\pi^-\pi^+\pi^-\nu_{\tau}$	& 
$K^-\pi^+\pi^-\nu_{\tau}$	& $K^-K^+\pi^-\nu_{\tau}$ & 
$K^-K^+K^-\nu_{\tau}$ \\
\hline
{\pipipi} & $0.22~(-0.1\%)$ 			& $0.079~(-0.1\%)$	 	& $0.022~(-1.1\%)$	 	& $6.4\times 10^{-3}~(+2.6\%) $  \\
{\Kpipi}  & $0.012~(-0.2\%)$			& $0.18~(-0.5\%)$		& $0.047~(+1.4\%)$		& $0.019~(-3.1\%)$	\\
{\KKpi}   & $3.9\times 10^{-4}~(-0.3\%)$	& $4.7\times 10^{-3}~(-4.4\%)$ 	& $0.12~(-1.0\%)$		& $0.051~(-0.4\%)$	\\
{\KKK}    & $5.0\times 10^{-6}~(+4.5\%)$	& $1.3\times 10^{-4}~(-9.1\%)$	& $2.3\times 10^{-3}~(-14.0\%)$	& $0.081~(+1.2\%)$	\\
\hline
\end{tabular}
\label{EfficiencyTable}
\end{table}
\end{center}

\section{Branching fraction and mass spectrum calculation}
\label{Section:CalculationOfBranchingRatio}

The numbers of candidate events in the  
{\pipipi}, {\Kpipi}, {\KKpi}, and {\KKK} modes, after applying 
all selection criteria, are summarized in Table~\ref{NumberOfEvents1}. 
Possible backgrounds in these signal candidate samples include
(a) cross-feed from the signal modes and
(b) other processes such as {\pipipipizero}, {\Kspi}, 
as well as continuum {\eeqq}.
The fractions of backgrounds coming from other processes are
5\% to 12\%, as summarized in Table~\ref{NumberOfEvents1} 
(fourth column) for each signal mode.
The fifth column shows the fraction of the main  background component as determined from MC.
The mode {\pipipipizero} dominates 
for the {\pipipi} and {\Kpipi} modes,
while the background from continuum {\eeqq} is 
dominant for the {\KKpi} and {\KKK} modes.
\begin{center}
\begin{table}[htb]
\caption{The number of reconstructed events (second column), 
the number of true events (third column),
the number of background events other than three-prong cross-feeds 
(fourth column), and the main source of other backgrounds with 
its fraction of the total (fifth column).}
\begin{tabular}{l|c|c|c|l}
\hline
Decay mode & $N^{\rm {rec}}$ & $N^{\rm{true}}$ & $N^{\rm {other}}$ & 
Main component in $N^{\rm {other}}$ \\
\hline
{\pipipi} & $8.86\times10^6$	& $3.52\times10^7$ 	
	& $9.35\times10^5~(10.6\%)$ 	& {\pipipipizero} (64.2\%)	\\
{\Kpipi} & $7.94\times10^5$ 	& $1.38\times10^6$
	& $9.60\times10^4~(12.1\%)$	& {\pipipipizero} (34.4\%)	\\
{\KKpi} & $1.08\times10^5$	& $6.47\times10^5$
	& $7.16\times10^3~(6.66\%)$	& {\eeqq} (30.3\%)	\\
{\KKK} 	& $3.16\times10^3$	& $1.37\times10^4$
	& $1.71\times10^2~(5.41\%)$	& {\eeqq} (53.0\%)	\\
\hline
\end{tabular}
\label{NumberOfEvents1}
\end{table}
\end{center}

To take into account the cross-feed between the decay channels, 
the true yield 
$N_{i}^{\rm {true}}$ 
($i$ is the mode number and the values from 1 to 4 correspond to {\pipipi}, {\Kpipi}, {\KKpi}, and {\KKK}),
is obtained using the following equation:
\begin{equation}
N^{\rm {true}}_i = 
\sum_j (\mathcal{E}^{-1})_{ij} (N^{\rm {rec}}_j - N^{\rm {other}}_j)~,
\label{breq1}
\end{equation}
where $N_{j}^{\rm {rec}}$ is the number of events for the 
$j$-th reconstructed decay mode 
and $N_{j}^{\rm {other}}$ is the number of 
background events in the $j$-th mode.
Here $\mathcal{E}_{ij}$ is the efficiency for detecting the $j$-th
mode  as the $i$-th one.
The values of $N^{\rm {rec}}$ and $N^{\rm {other}}$
are listed in the second and third columns of Table~\ref{NumberOfEvents1}, respectively. 

We have determined the efficiency (migration) matrix
iteratively. First, it is
determined using the mass spectra as modeled in the current TAUOLA MC program.
(We refer to this efficiency matrix as $\mathcal{E}^{(1)}_{ij}$).
As is discussed in the next paragraph, we then unfolded the mass
spectra of each mode to obtain the true mass distribution. 
The resultant unfolded mass spectra are then used to determine the
efficiency matrix $\mathcal{E}^{(2)}_{ij}$ using a weighting procedure.
The matrix elements  $\mathcal{E}^{(2)}_{ij}$  are given in Table \ref{EfficiencyTable}, 
where the values given in parentheses are the percentage differences between
$\mathcal{E}^{(1)}_{ij}$ and $\mathcal{E}^{(2)}_{ij}$;
$(\mathcal{E}^{(2)}_{ij} - \mathcal{E}^{(1)}_{ij})/\mathcal{E}^{(1)}_{ij}$.
Since the efficiency is smooth as a function of hadronic mass, the
difference between $\mathcal{E}^{(1)}_{ij}$ and $\mathcal{E}^{(2)}_{ij}$ is very small ($\sim$1\%)
in most cases, except for the off-diagonal entries related to the {\KKK} mode.
The source of these large discrepancies is the mass dependence of the fake rates, 
together with the considerable differences between the unfolded spectrum and the generated spectrum. 

The unfolding procedure extracts the true mass spectra
for all four modes, denoted by the vector $\boldsymbol x$,
by solving the matrix equation 
$\hat{A} \boldsymbol{x} = \boldsymbol{b}$.
In this equation, $\boldsymbol b$ is the vector of the reconstructed mass spectrum 
with backgrounds other than the 
three-prong cross-feed subtracted,
and $\hat{A}$ is 
the response matrix.

\begin{figure}[!htbp]
\begin{center}
\includegraphics[width=0.9\textwidth]
{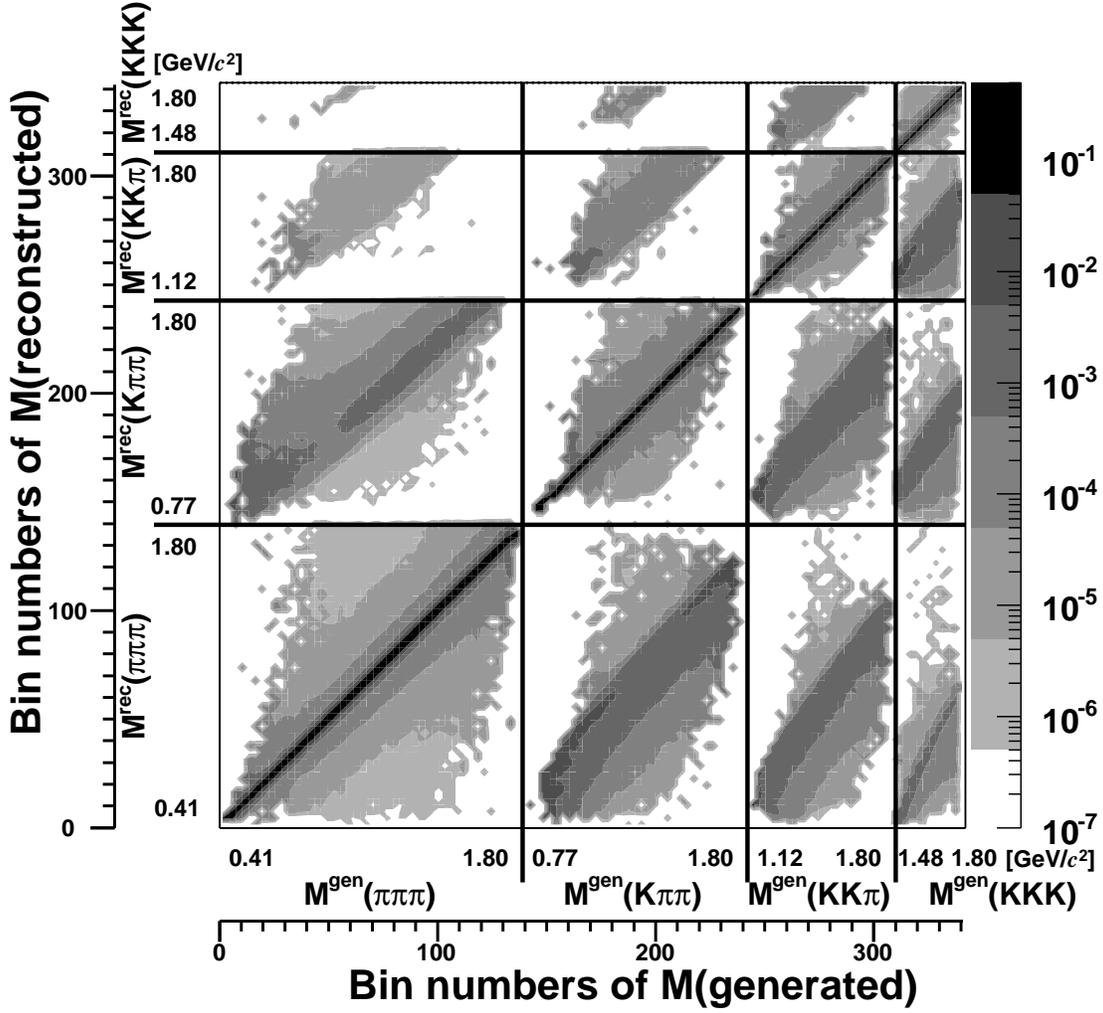}
\caption{Response matrix $\hat{A}$ obtained by merging 
16 sub-response matrices that represent the efficiency or 
fake rates among the four three-prong modes. 
The bin numbers correspond to the following ranges of the invariant masses:
(1) from 0 to 138 , 0.41 {\GeVcc} to 1.8 {\GeVcc} of {\Mpipipi}; 
(2) from 139 to 241 , 0.77 {\GeVcc} to 1.8 {\GeVcc} of {\MKpipi}; 
(3) from 242 to 309 , 1.12 {\GeVcc} to 1.8 {\GeVcc} of {\MKKpi}; 
and (4) from 310 to 341 , 1.48 {\GeVcc} to 1.8 {\GeVcc} of {\MKKK}. 
All bin sizes are 10 {\MeVcc}.
}
\label{Unfolding_Ahat}
\end{center}
\end{figure}

The matrix $\hat{A}$ takes into account
the cross-feed between different modes
caused by particle misidentification, as well as the effects of
finite resolution and the limited acceptance
of the detector. The response matrix is obtained by merging 16
sub-response matrices as shown in Fig.~\ref{Unfolding_Ahat}.
The matrix is determined using the TAUOLA decay model MC simulation.
Clear correlations between the measured and generated values are
seen along the diagonal parts of $\hat{A}$.

\begin{figure}[htb]
\begin{center}
\includegraphics[width=0.9\textwidth]
{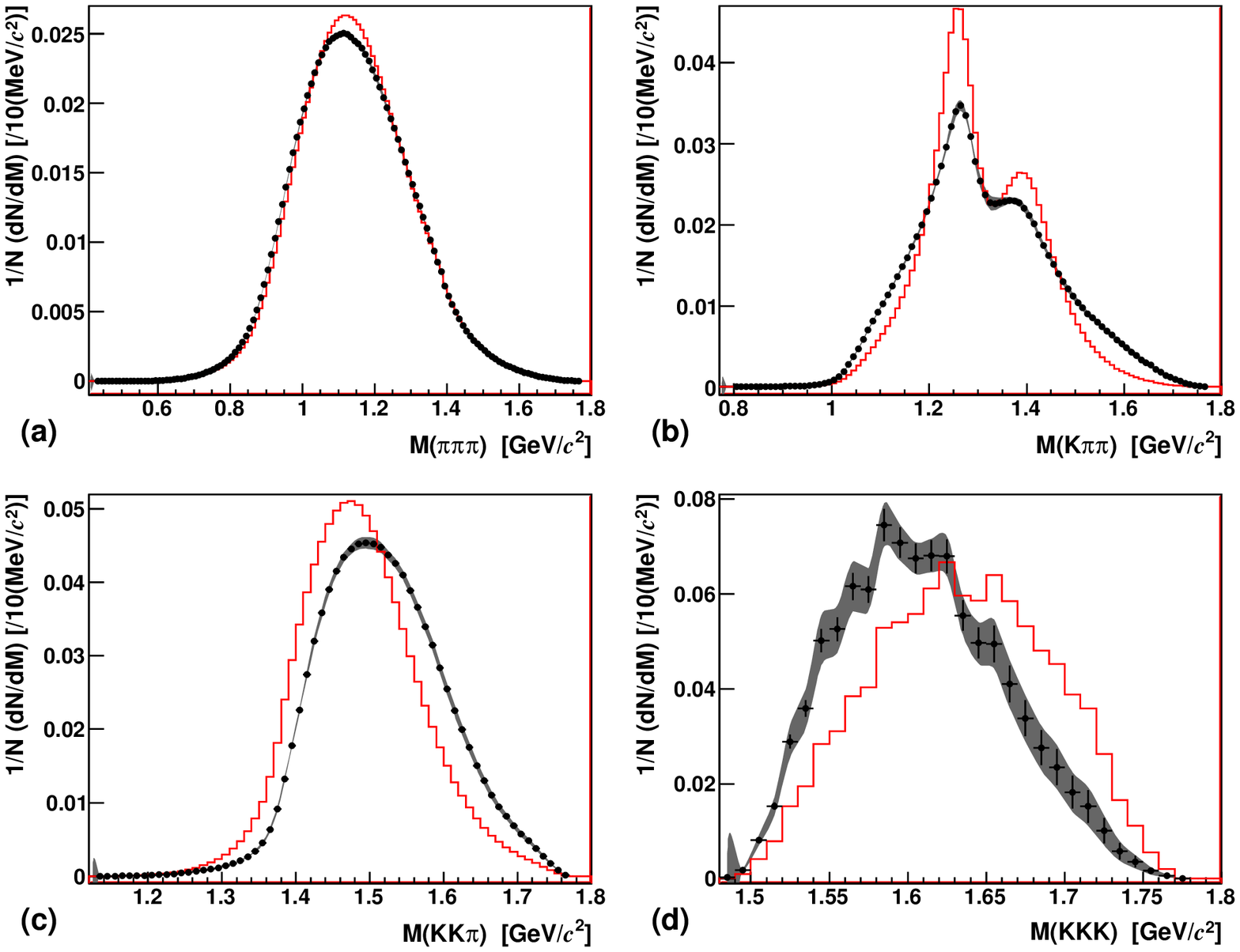}
\caption{The unfolded mass spectra of the three-prong decays, $1/N (dN/dM)$: 
(a) {\Mpipipi} distribution of {\pipipi}, 
(b) {\MKpipi} distribution of {\Kpipi}, 
(c) {\MKKpi} distribution of {\KKpi}, and 
(d) {\MKKK} distribution of {\KKK}. 
The black points correspond to the unfolded mass spectra with statistical uncertainties only, and
the gray bands correspond to the systematic uncertainty.
The solid histograms are the model predictions
implemented in the current TAUOLA MC simulation.
}
\label{Unfolding_Minv}
\end{center}
\end{figure}

For the unfolding algorithm, 
we follow the method of the ALEPH Collaboration, 
which is based on the Singular Value 
Decomposition technique 
for matrix inversion~\cite{Citation:Unfolding_Hoecker}   
and advanced regularization~\cite{Citation:Tikhonov}. 
Two independent {\hhh} signal MC samples are used to check
the validity of the unfolding procedure.  
One sample is used for the determination of the response
matrix. 
We check the reproducibility of the generated $M(hhh)$ mass
spectra with the other sample, and find, for the whole mass range, that the
differences are less than 1\%.
These discrepancies are included in the systematic uncertainties of unfolding 
(UNF1 below).
The resulting unfolded 
normalized
mass spectra of 
the three-prong decays are shown in Fig. \ref{Unfolding_Minv},
where the error bars indicate the statistical uncertainty
only, and the gray bands correspond to the systematic uncertainty.
Note that the uncertainties common for all bins are taken into
account as the systematic uncertainties on the branching fraction (Table~\ref{Systematics}), 
while the bin-by-bin errors remained in the normalized mass spectra $1/N (dN/dM)$ are accounted here.
The following sources of systematic uncertainties in the mass spectra
are considered: 
the unfolding procedure (UNF1 and UNF2),
the kaon identification efficiency (KID), 
the background estimation other than the three-prong cross-feed (BGE),  
the effect of $\gamma$ veto (GAM), and
the track momentum scale (MOM).
Table~\ref{Minv_uncertainties} summarizes various contributions to 
the systematic uncertainties on the normalized mass spectra $1/N (dN/dM)$ for each decay mode. 

The systematic uncertainty due to the unfolding procedure is determined from MC 
by comparing the generated and unfolded mass spectra (UNF1). Another estimation of the 
uncertainty of the unfolding procedure is obtained by changing the value of unfolding parameter
that determines the effective rank of the response matrix (UNF2).
This is the most important systematic uncertainty for $KKK$ mode.
The kaon identification efficiency and the fake rate are
calibrated by using $D^{*+}$ sample as described in the previous section. 
The uncertainties due to the particle identification (KID)
are evaluated by varying efficiency and cross-feed fraction by one 
standard deviation in the response matrix.
The systematic uncertainty from the background estimation 
(BGE) is evaluated  
by varying  the branching fraction values used in the MC by one standard deviation.
The uncertainty due to the $\gamma$ veto (GAM) is evaluated by using different 
selection criteria for photon energy 
and comparing the resultant unfolded spectra.
For the uncertainty from the track momentum scale (MOM), 
we reconstruct $\phi \rightarrow K^+ K^-$ and $D^\pm \rightarrow K^\mp \pi^\pm \pi^\pm$ decays.
By comparing the reconstructed masses of $\phi$ and $D^\pm$ mesons with their world average values, 
we conservatively assign 0.01\% 
uncertainty for the momentum scale. 
By applying this variation to the reconstructed mass spectra of {\hhh} decay and unfolding it, 
we obtain the uncertainties due to the momentum scale. 

\begin{figure}[htb]
\begin{center}
\includegraphics[width=0.8\textwidth]
{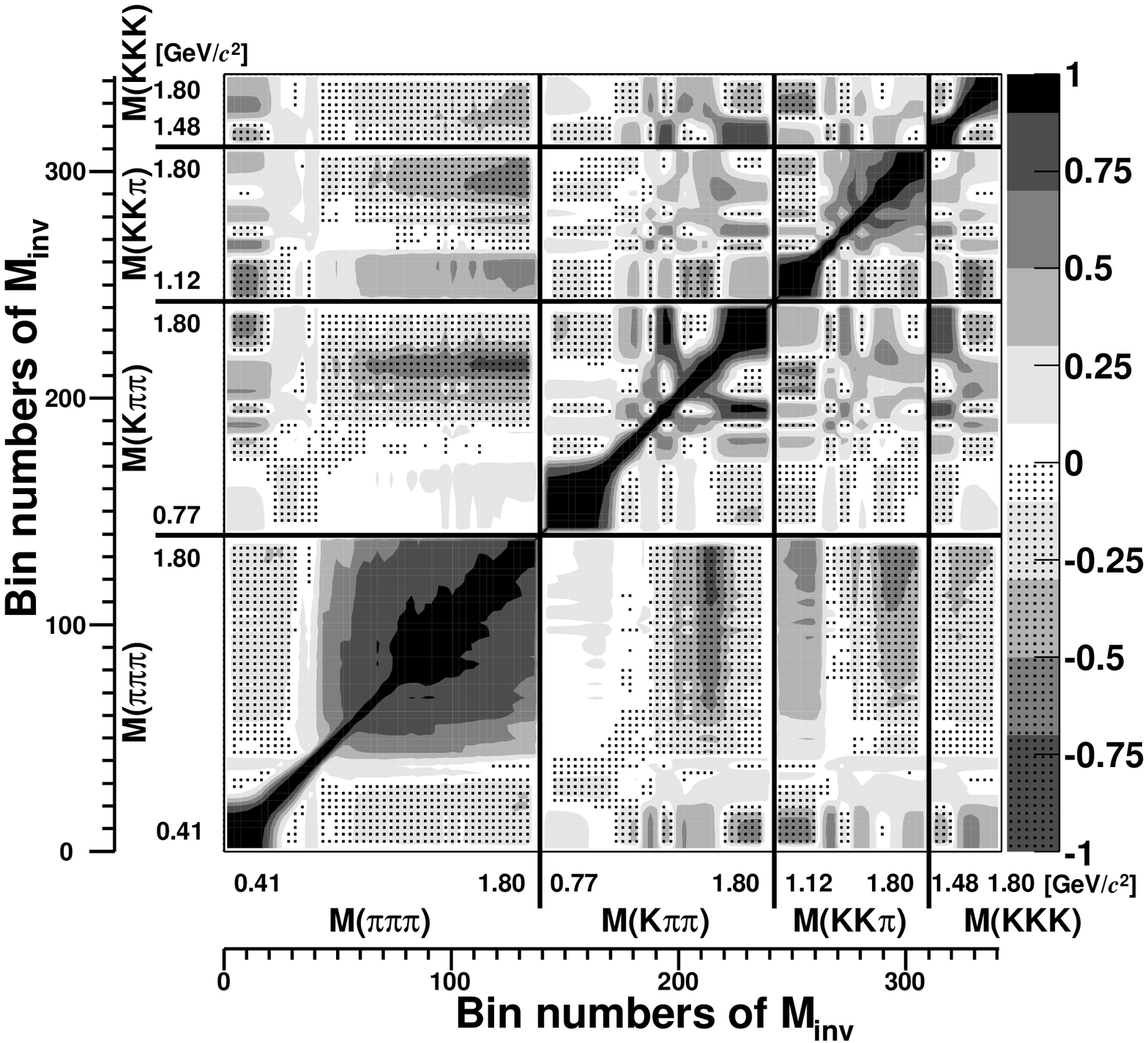}
\caption{
The correlation coefficients of the unfolded mass spectra for three-prong decays. 
See Fig.~\ref{Unfolding_Ahat} for the convention of the bin numbers.
The pattern is the reflection of the detailed unfolding procedure.
}
\label{Unfolding_corr}
\end{center}
\end{figure}

The systematic uncertainty varies as a function 
of the mass of the hadronic system in each mode. 
The typical uncertainties averaged over all mass bins
are evaluated by summing the systematic uncertainties of each bin and ignoring the correlations between bins, 
and the results are 0.7\%, 2.2\%, 2.2\%, and 9.5\%, for the {\pipipi}, 
{\Kpipi}, {\KKpi}, and {\KKK} decay modes, respectively.  
In practice, we have obtained the covariance matrix for
each source of the systematic uncertainties
and added them to obtain the total uncertainty.
The gray bands in Fig.~ \ref{Unfolding_Minv}
show the square roots of the diagonal components of the covariance matrix.
The off-diagonal components of the covariance matrix are shown 
in Fig.~\ref{Unfolding_corr}
in term of the correlation coefficients
$\rho^{\rm{Spec}}(i,j)$
defined by
$\rho^{\rm{Spec}}(i,j)= {\rm{cov}}(i,j)/\sqrt{{\rm{cov}}(i,i) \cdot {\rm{cov}}(j,j)}$,
where $i$ and $j$ are bin numbers.
%
\begin{center}
\begin{table}[htb]
\caption{
Summary of the relative errors of the unfolded mass spectra, $1/N (dN/dM)$ (in \%) from different sources of uncertainties: the unfolding procedure (UNF1, UNF2), the kaon identification (KID), the background estimation (BGE), the $\gamma$ veto (GAM), and the momentum scale (MOM). The ``average'' uncertainties (the 2nd, 4-th, 6-th, and 8-th columns) are evaluated by 
taking average of errors in all bins.
The ``peak'' uncertainties (the 3rd, 5-th, 7-th, and 9-th columns) represent the errors at the peak position of the spectra.
See the text for a more detailed description.
}
\begin{tabular}{l|c|c|c|c|c|c|c|c}
\hline
\hline
Sources & \multicolumn{2}{c|}{$\tau \rightarrow$} & \multicolumn{2}{c|}{$\tau \rightarrow$} & \multicolumn{2}{c|}{$\tau \rightarrow$} & \multicolumn{2}{c}{$\tau \rightarrow$}         \\
of  & \multicolumn{2}{c|}{$\pi\pi\pi\nu$}     & \multicolumn{2}{c|}{$K\pi\pi\nu$}       & \multicolumn{2}{c|}{$KK\pi\nu$}         & \multicolumn{2}{c}{$KKK\nu$}           \\
\cline{2-9}
uncertainties & average & peak & average & peak & average & peak & average & peak \\
\hline
(UNF1) 		  & 0.5 & 0.5 & 0.1 & 0.1 & 0.4 & 0.4 & 0.6 & 0.6 \\
(UNF2) 		  & 0.1 & 0.1 & 1.6 & 1.5 & 0.9 & 0.9 & 7.3 & 4.0\\
(KID) 		  & 0.4 & 0.4 & 1.0 & 0.8 & 1.5 & 1.1 & 1.9 & 0.9\\
(BGE) 		  & 0.2 & 0.2 & 0.6 & 0.5 & 0.4 & 0.3 & 1.9 & 1.1\\
(GAM) 		  & 0.6 & 0.4 & 0.9 & 0.4 & 1.2 & 0.9 & 4.6 & 3.2\\
(MOM) 		  & 0.1 & 0.0 & 0.4 & 0.2 & 0.3 & 0.3 & 2.7 & 3.1\\
\hline
Total systematics & 0.7 & 0.7 & 2.2 & 1.9 & 2.2 & 1.7 & 9.5 & 6.2\\
Statistics 	  & 0.2 & 0.2 & 1.1 & 0.8 & 1.0 & 0.8 & 7.1 & 4.6\\
\hline
Total 		  & 0.9 & 0.8 & 2.4 & 2.0 & 2.4 & 1.9 & 11.8 & 7.7\\
\hline
\hline
\end{tabular}
\label{Minv_uncertainties}
\end{table}
\end{center}

Figure~\ref{Unfolding_Minv} shows rather large discrepancies 
between the mass spectra
of the generator and 
our unfolded spectra except for {\pipipi} decay.
For example, for the {\Kpipi} mass spectrum (Fig. \ref{Unfolding_Minv}b), 
it is evident that there are contributions 
from two resonant peaks,
similar to the MC expectations.
However, the faster rise of the real data as compared with the MC 
in the $1.0 - 1.1$ GeV mass region and the fall-off 
at high masses in the {\Kpipi} mass spectrum 
may also be suggestive of a substantial non-resonant component 
that is not included in the current MC model. 
The use of the unfolded
invariant mass distributions 
in estimating the 
efficiencies
allows one to compensate for   
model deficiencies  
when evaluating the branching fractions. 

In order to determine the branching fraction, we normalize to the number 
of pure leptonic decays, where one $\tau$ decays to {\enunu} 
and the other decays to {\mununu}~\cite{Citation:Epifanov} 
(hereafter such events are referred to as {\emuevent} events).
The branching fraction for the $i$-th decay mode can then 
be written as:
\begin{equation}
\mathcal{B}_i 
  = N^{\rm {true}}_{i} \cdot \frac{ \varepsilon_{e\mu} }
{N_{\rm {sig},e\mu}}   \cdot 
\frac{\mathcal{B}_{\tau\rightarrow e\overline{\nu}\nu}\cdot 
\mathcal{B}_{\tau\rightarrow\mu\overline{\nu}\nu}}
{\mathcal{B}_{\tau\rightarrow l\overline{\nu}\nu}}~, 
\end{equation}
where  
$N_{ \rm{sig},e\mu} = 8.1\times10^6$ and $\varepsilon_{e\mu} = 0.221 \pm 0.003$ are the number of 
{\emuevent} events and the corresponding detection efficiency, respectively. 
The uncertainty in $\varepsilon_{e\mu}$ is determined from the lepton identification uncertainties. 
Here $\mathcal{B}_{\tau\rightarrow e\overline{\nu}\nu} = 0.178$ and 
$\mathcal{B}_{\tau\rightarrow\mu\overline{\nu}\nu} = 0.174$ are the branching 
fractions of {\enunu} and {\mununu} decays, respectively, 
and $\mathcal{B}_{\tau\rightarrow l\overline{\nu}\nu} = 
\mathcal{B}_{\tau\rightarrow e\overline{\nu}\nu} + 
\mathcal{B}_{\tau\rightarrow\mu\overline{\nu}\nu}$.
The true yields $N_{i}^{\rm {true}}$ obtained from Eq.~\ref{breq1}
are given in the third column of Table~\ref{NumberOfEvents1}. 

\begin{figure}[htb]
\begin{center}
\includegraphics[width=0.9\textwidth]
{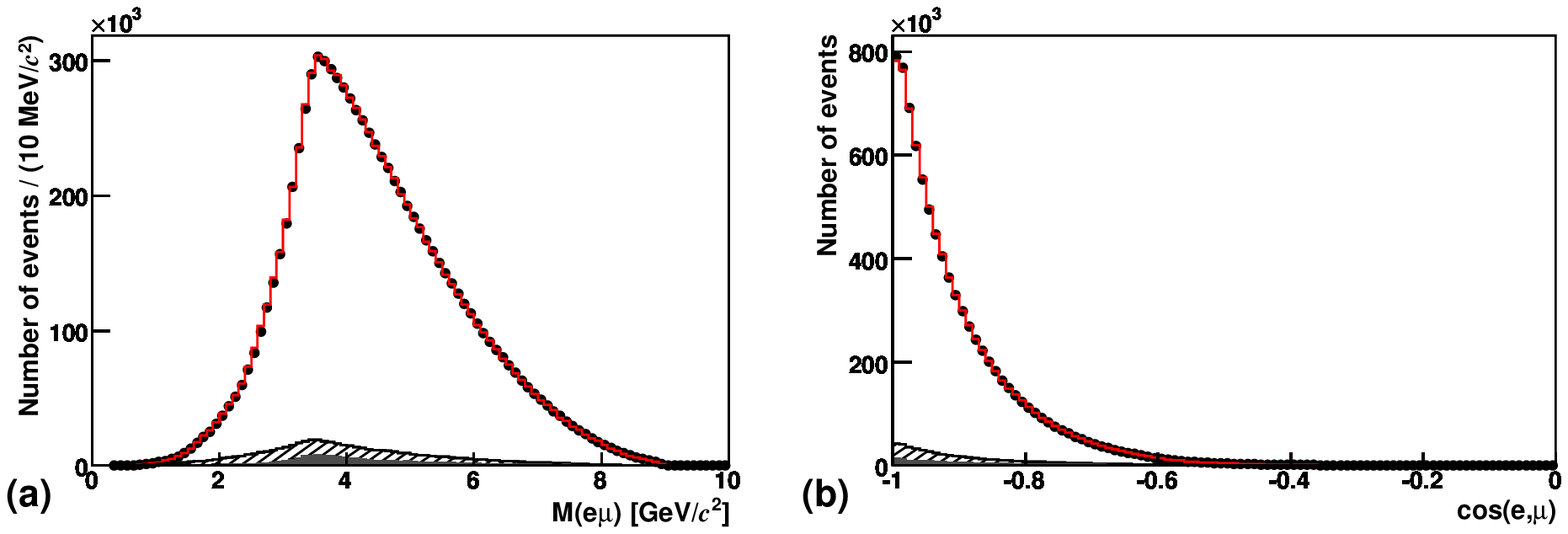}
\caption[{\emuevent} event selection (M(e$\mu$), \cos(e,$\mu$))]
{Distributions of the {\emuevent} events: (a) Invariant mass of the electron--muon system. 
(b) Cosine of the angle between the electron and muon. 
The closed circles and the solid histogram represent the data and 
the sum of the MC expectations, respectively. The open histogram represents 
the {\emuevent} signal events, 
while the $\tau$-pair background 
and the two-photon background are shown 
by the hatched and the gray histograms, respectively. }
\label{EmuSelection1}
\end{center}
\end{figure}

This normalization method requires a precise measurement of  
the number of {\emuevent} events and a careful determination of the corresponding efficiency. 
Figure~\ref{EmuSelection1} shows a comparison of the invariant mass 
of the electron and muon system and the cosine of the angle 
between electron and muon, for real {\emuevent} events and the 
sum of MC expectations. The MC reproduces the data reasonably well. 
The background fraction is estimated to be $5.9\%$ using MC.
After subtracting background,
we obtain the number of {\emuevent} events. 

\begin{figure}[htb]
\begin{center}
\includegraphics[width=0.9\textwidth]
{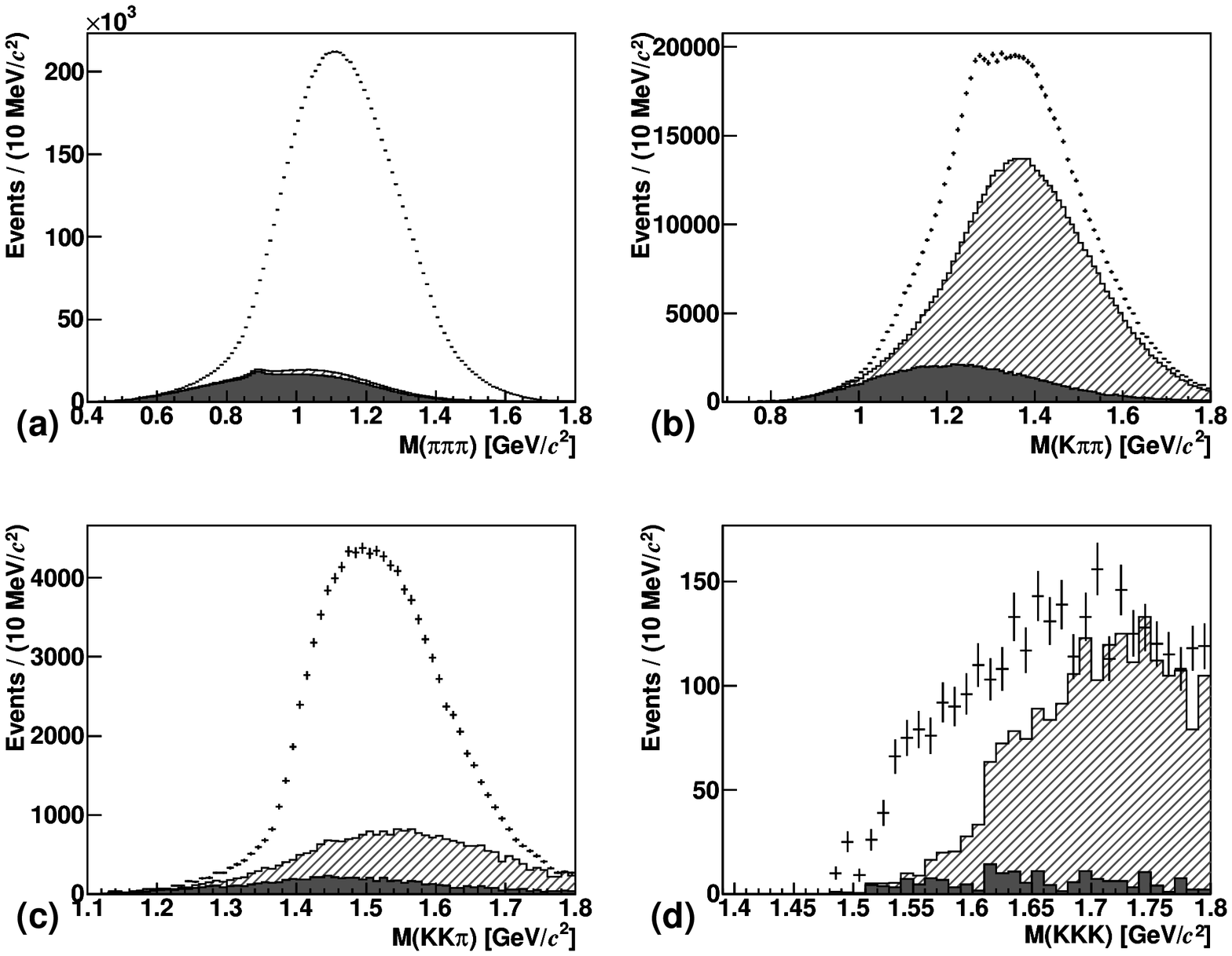}
\caption{The reconstructed invariant mass distributions: 
(a) {\Mpipipi} distribution for {\pipipi}, 
(b) {\MKpipi} distribution for {\Kpipi}, 
(c) {\MKKpi} distribution for {\KKpi}, and 
(d) {\MKKK} distribution for {\KKK}. 
For all histograms, the black 
points with error bars represent the data, the gray histograms are the sum of 
all backgrounds other than the three-prong cross-feed, 
and the hatched histograms are the sum of three-prong cross-feed backgrounds. 
Note that the cross-feed backgrounds are estimated from the unfolding of the data.
The uncertainties shown for the data are statistical only.}
\label{ResultBeforeUnfolding}
\end{center}
\end{figure}

Figure~\ref{ResultBeforeUnfolding} shows the
observed distributions of {\Mpipipi}, {\MKpipi}, {\MKKpi}, and {\MKKK}.
The backgrounds are represented by the hatched and the gray histograms
for the sum of the three-prong cross-feed backgrounds and other background components, respectively.
Note that the three-prong cross-feed backgrounds 
are determined by the unfolding procedure previously described.

Table~\ref{Systematics} summarizes various contributions to 
the systematic uncertainties on the branching fractions.
The uncertainty in the track-finding efficiency 
is estimated from a comparison of real and MC data 
for the $D^* \rightarrow \pi D^0$,
$D^0 \rightarrow \pi\pi K_S^0(K_S^0 \rightarrow \pi^+ \pi^-)$ decay sample. 
The uncertainties due to the kaon and pion identification are evaluated 
by applying the uncertainties in kaon identification efficiencies to the efficiency matrix in Table \ref{EfficiencyTable}.
First, we measure the kaon and pion identification efficiencies and their uncertainties 
using the control sample of $D^{*+} \rightarrow D^0 \pi^+_s$ 
and $D^0 \rightarrow K^- \pi^+$ events. 
Since these efficiencies are applied to the 
efficiency matrix,
we propagate the uncertainties in the kaon and pion identification efficiencies
to the covariance matrix of branching fraction measurement.
The kaon identification systematic uncertainties for {\pipipi}, {\Kpipi}, {\KKpi}, and {\KKK} decays are 
1.3\%, 3.9\%, 1.9\%, and 5.4\%, respectively.
For lepton identification efficiency uncertainties, 
we use $\gamma \gamma \rightarrow e^+ e^- / \mu^+ \mu^-$ events. 

In order to take into account the effects of the mass spectra on
the branching fraction measurements, 
we have determined the mass spectra of the four dominant
decay modes iteratively. 
The uncertainties in branching fractions due to the mass spectra are evaluated by varying the
unfolded spectra by one
standard deviation of their statistical and systematic errors, 
assuming 100\% bin-by-bin correlations for systematic errors. 
%
The trigger efficiency for the three-prong modes 
is $\sim$86\% with an uncertainty of $\sim$0.6\%. 
This efficiency is obtained using a trigger simulation program applied to the signal MC samples. 
The uncertainty due to the $\gamma$ veto is evaluated by using 
different selection criteria for photon energy. 
The uncertainty due to the {\Ks} veto is evaluated by removing 
the veto requirement. 
The uncertainties in background estimation include the effect of 
luminosity, the cross section for {\eetautau}, and the 
subtraction of background from non-three-prong $\tau$ decays and continuum {\eeqq}.
The uncertainty on the $\tau$ decay background subtraction is evaluated 
by propagating the errors from the $\tau$ branching fractions
other than those with three-prongs. 
The fraction of the {\eeqq}  background is negligible for
{\pipipi} and {\Kpipi} modes, and is
$2.0\pm0.2\%$ and $2.9\pm0.3\%$ for {\KKpi} and {\KKK} modes, respectively.
The uncertainties in the {\eeqq} background are estimated from comparison
of the number of events in data and {\eeqq} MC,
in the mass region above the $\tau$ mass.
The effects of the uncertainties in the luminosity and the cross section for
{\eetautau} \cite{Citation:TauTauCrosssection}  
are negligible 
because we use {\emuevent} events for normalization. 
The uncertainties in the leptonic decay branching fractions for the
$\tau$ are also taken into account. 

\begin{center}
\begin{table}[htp]
\caption{Summary of the systematic uncertainties 
on the branching fractions
(\%).}
\begin{tabular}{l|c|c|c|c}
\hline
 & $\tau \rightarrow$	& $\tau \rightarrow$	& $\tau \rightarrow$ & $\tau \rightarrow$ 	\\
Source & $\pi\pi\pi\nu$	& $K\pi\pi\nu$	& $KK\pi\nu$	& $KKK\nu$ \\ \hline
Tracking efficiency & $+2.2$/$-2.0$ & $+2.1$/$-2.0$ & $+2.1$/$-2.0$ & $+2.1$/$-1.9$ \\
Particle ID &
$\pm1.9$ & $+4.0$/$-4.1$ & $\pm2.3$ & $+5.4$/$-5.6$ \\
Mass spectrum &
$\pm0.1$ & $\pm0.1$ & $\pm0.1$ & $+0.5$/$-0.8$ \\
Trigger efficiency & $\pm0.5$ & $\pm0.5$ & $\pm0.6$ & $\pm0.6$ \\
$\gamma$ veto & $\pm0.9$ & $\pm1.5$ & $\pm1.7$ & $\pm0.5$ \\
{\Ks} veto & $\pm0.2$ & $\pm0.2$ & $\pm0.1$ & $\pm0.3$ \\
Background estimation & 
$\pm0.3$ & $\pm1.3$ & $\pm0.3$ & $\pm0.4$ \\
Leptonic branching fraction & $\pm0.2$ & $\pm0.2$ & $\pm0.2$ & $\pm0.2$ \\
\hline
Total & 
$+3.1$/$-3.0$ & $\pm5.0$ & $\pm3.6$ & $+5.9$/$-6.0$ \\
\hline
\end{tabular}
\label{Systematics}
\end{table}
\end{center}

After taking into account the backgrounds, the efficiencies and 
various sources of systematic uncertainties discussed above, 
we obtain the following branching fractions for three-prong decays:

\begin{eqnarray}
\mathcal{B}(\tau^-\rightarrow \pi^-\pi^+\pi^-\nu_{\tau}) & = &	
(8.42 \pm 0.00(\rm stat.) ^{+0.26} _{-0.25} (\rm sys.)) \times 10^{-2}~, \nonumber  \footnotemark \\
\mathcal{B}(\tau^-\rightarrow K^- \pi^+\pi^-\nu_{\tau}) & = &
(3.30 \pm 0.01(\rm stat.) ^{+0.16} _{-0.17} (\rm sys.)) \times 10^{-3}~, \nonumber  \\
\mathcal{B}(\tau^-\rightarrow K^-K^+\pi^-\nu_{\tau}) & = &
(1.55 \pm 0.01(\rm stat.) ^{+0.06} _{-0.05} (\rm sys.)) \times 10^{-3}~,\textrm{~and}  
\nonumber  \\
\mathcal{B}(\tau^-\rightarrow K^-K^+K^-\nu_{\tau}) & = &	
(3.29 \pm 0.17(\rm stat.) ^{+0.19} _{-0.20} (\rm sys.)) \times 10^{-5}~. \nonumber 
\end{eqnarray} 
\footnotetext{
The actual value of the statistical uncertainty of $\mathcal{B}(\tau^-\rightarrow \pi^-\pi^+\pi^-\nu_{\tau})$ is $0.003\times10^{-2}$.
} 

The covariance matrices for the branching fraction measurements are 
given in Table \ref{Correlation} in terms of the correlation coefficients
$\rho^{\rm{BF}}(i,j) = {\rm{cov}}(i,j)/\sqrt{{\rm{cov}}(i,i) \cdot {\rm{cov}}(j,j)}$,
where $i$ and $j$ correspond to one of the four three-prong decay modes.
There is a strong correlation between {\pipipi} and {\Kpipi} decays, 
which mostly comes from 
the effect of kaon identification.

\begin{center}
\begin{table}[htp]
\caption{The correlation coefficients in the branching fraction measurements.}
\begin{tabular}{l|ccc}
\hline
 & $\tau^-\rightarrow$                  & $\tau^-\rightarrow$           &  $\tau^-\rightarrow$ \\
 & $K^-\pi^+\pi^-\nu_{\tau}$    &$K^-K^+\pi^-\nu_{\tau}$        & $K^-K^+K^-\nu_{\tau}$ \\
\hline
{\pipipi} & $+0.175$ & $+0.049$ & $-0.053$ \\
{\Kpipi}  &          & $+0.080$ & $+0.035$ \\
{\KKpi}   &          &          & $-0.008$ \\
\hline
\end{tabular}
\label{Correlation}
\end{table}
\end{center}

Using the results of the branching fractions and the correlation matrix, the 
ratios of the branching fractions involving kaons
to the branching fraction of {\pipipi} decay are estimated:
\begin{eqnarray}
\mathcal{B}(\tau^-\rightarrow K^- \pi^+\pi^-\nu_{\tau}) / \mathcal{B}(\tau^-\rightarrow \pi^-\pi^+\pi^-\nu_{\tau}) & = &	
(3.92 \pm 0.02 ^{+0.15} _{-0.16}) \times 10^{-2}~, \nonumber  \\
\mathcal{B}(\tau^-\rightarrow K^-K^+\pi^-\nu_{\tau}) / \mathcal{B}(\tau^-\rightarrow \pi^-\pi^+\pi^-\nu_{\tau}) & = &	
(1.84 \pm 0.01 \pm 0.05) \times 10^{-2}~,\textrm{~and}  \nonumber  \\
\mathcal{B}(\tau^-\rightarrow K^-K^+K^-\nu_{\tau}) / \mathcal{B}(\tau^-\rightarrow \pi^-\pi^+\pi^-\nu_{\tau}) & = &	
(3.90 \pm 0.20 ^{+0.22} _{-0.23}) \times 10^{-4}~, \nonumber 
\end{eqnarray}
where the first and the second uncertainties are statistical and systematic, respectively.

\section{Discussion}

The results of this analysis for the branching fractions of various
three-prong modes are listed in Table~\ref{BRcompare} together with
recent results from BaBar~\cite{Citation:BABAR}. 
Note that 
for the {\pipipi} and {\Kpipi} modes, the branching fractions listed 
do not include any $K^0$ contribution.
%
\begin{center}
\begin{table}[htp]
\caption{
Comparison of the branching fraction results
}
\begin{tabular}{l|c|c}
\hline
Decay mode & BaBar & Belle \\
\hline
{\pipipi}, \%	& $8.83 \pm 0.01 \pm 0.13$	& 
$8.42 \pm 0.00 ^{+0.26} _{-0.25}$	\\ 
{\Kpipi}, \%	& $0.273 \pm 0.002 \pm 0.009$	& 
$0.330 \pm 0.001 ^{+0.016} _{-0.017}$		\\ 
{\KKpi}, \%	& $0.1346 \pm 0.0010 \pm 0.0036$ & 
$0.155 \pm 0.001 ^{+0.006} _{-0.005}$		\\ 
{\KKK}, $10^{-5}$ & $1.58 \pm 0.13 \pm 0.12$	& 
$3.29 \pm 0.17 ^{+0.19} _{-0.20}$		\\ 
\hline
\end{tabular}
\label{BRcompare}
\end{table}
\end{center}

\begin{figure}[htb]
\begin{center}
\includegraphics[width=0.9\textwidth]
{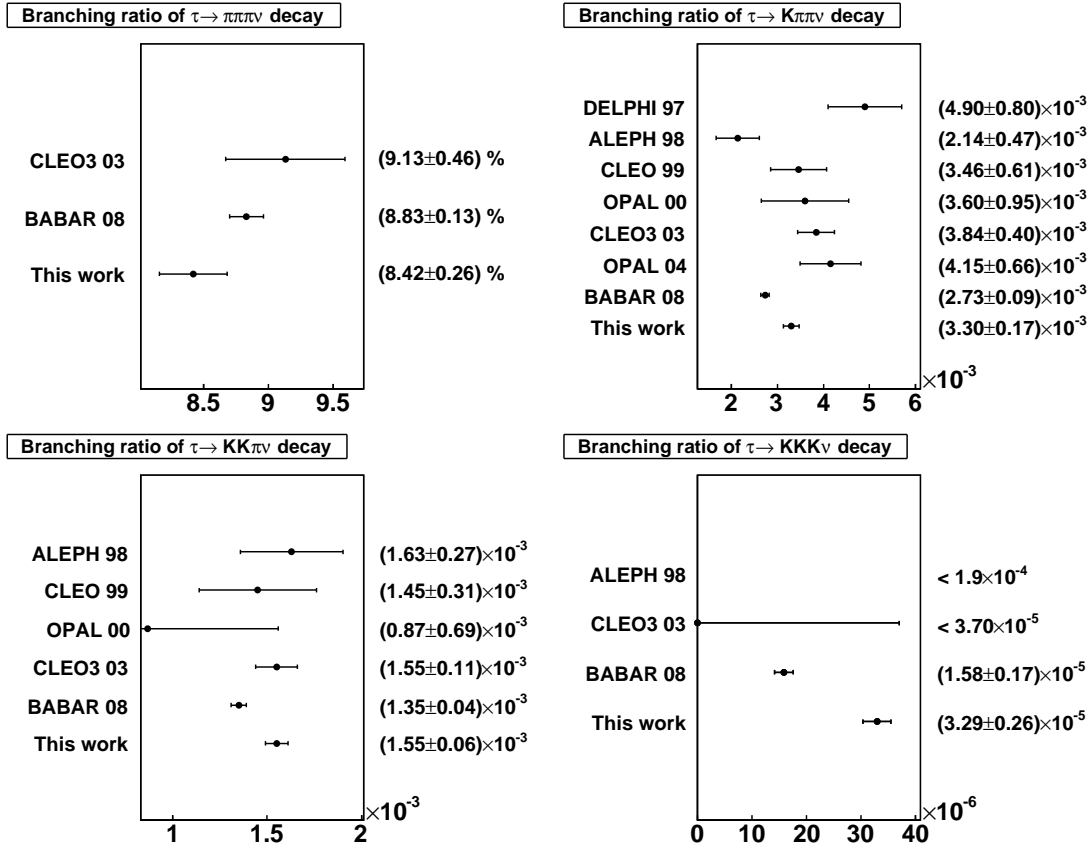}
\caption{
Summary of the branching fraction measurements of three-prong $\tau$ decays. 
}
\label{KpipiBranchingRatioHistory}
\end{center}
\end{figure}

In Fig.~\ref{KpipiBranchingRatioHistory}, our results are compared with the previous 
measurements. For all modes studied, the precision of the branching fractions
for both BaBar and Belle is significantly higher than before.
The accuracy of our results is comparable to that of BaBar, but the 
central values show striking differences in all channels other than 
{\pipipi}. For the {\pipipi} mode, our result is $1.4\sigma$
lower than that of BaBar, while for other modes the branching fractions
obtained by Belle are higher by $3.0\sigma$, $3.0\sigma$, and 
$5.4\sigma$ than those of BaBar for the {\Kpipi}, {\KKpi}, 
and {\KKK} modes, respectively.
 
\section{Conclusion}
Using a data sample of  $6.12 \times 10^8$ $\tau^+\tau^-$ pairs
collected with the Belle detector,
we measure the branching fractions for the {\pipipi}, {\Kpipi}, {\KKpi}, 
and {\KKK}  decay modes. 
We have been able to 
extract unfolded invariant mass
spectra for all four modes (Fig. \ref{Unfolding_Minv})
by taking into account the 
cross-feed effects.  
A future detailed analysis of these spectra as
well as those of the intermediate two-body states  will allow studies of the decay
dynamics to be performed. 
  
\section{Acknowledgments}
We thank the KEKB group for the excellent operation of the
accelerator, the KEK cryogenics group for the efficient
operation of the solenoid, and the KEK computer group and
the National Institute of Informatics for valuable computing
and SINET3 network support.  We acknowledge support from
the Ministry of Education, Culture, Sports, Science, and
Technology (MEXT) of Japan, the Japan Society for the 
Promotion of Science (JSPS), and the Tau-Lepton Physics 
Research Center of Nagoya University; 
the Australian Research Council and the Australian 
Department of Industry, Innovation, Science and Research;
the National Natural Science Foundation of China under
contract No.~10575109, 10775142, 10875115 and 10825524; 
the Department of Science and Technology of India; 
the BK21 and WCU program (grant No. R32-2008-000-10155-0) of the Ministry Education Science and
Technology, the CHEP SRC program and Basic Research program (grant No.
R01-2008-000-10477-0) of the Korea Science and Engineering Foundation,
Korea Research Foundation (KRF-2008-313-C00177),
and the Korea Institute of Science and Technology Information;
the Polish Ministry of Science and Higher Education;
the Ministry of Education and Science of the Russian
Federation and the Russian Federal Agency for Atomic Energy;
the Slovenian Research Agency;  the Swiss
National Science Foundation; the National Science Council
and the Ministry of Education of Taiwan; and the U.S.\
Department of Energy.
This work is supported by a Grant-in-Aid from MEXT for 
Science Research in a Priority Area ("New Development of 
Flavor Physics"), and from JSPS for Creative Scientific 
Research ("Evolution of Tau-lepton Physics").

\end{document}